\documentclass[12pt]{iopart}

\usepackage{graphicx}
\usepackage[usenames]{color}

\usepackage{iopams}  
\usepackage[colorlinks,linkcolor=red,anchorcolor=blue,
citecolor=blue]{hyperref}
\begin{document}

\title [Harmonically trapped attractive and repulsive SO and Rabi coupled 
BECs] 
{Harmonically trapped attractive and repulsive spin-orbit and Rabi coupled 
Bose-Einstein condensates}

\author{Emerson Chiquillo}
\address{Escuela de F\'isica, UPTC - Universidad Pedag\'ogica
y Tecnol\'ogica de Colombia,\\
Avenida Central del Norte, 150003 Tunja, Boyac\'a, Colombia}
\ead{emerson.chiquillo@uptc.edu.co}

\begin{abstract} 

Numerically we investigate the ground state of effective one-dimensional
spin-orbit (SO) and Rabi coupled two pseudo-spinor Bose-Einstein
condensates (BECs) under the effect of harmonic traps. 
For both signs of the interaction, density profiles of SO and Rabi coupled
BECs in harmonic potentials, which simulate a real experimental situation
are obtained.
The harmonic trap causes a strong reduction of the multi-peak nature of 
the condensate and it increases its density.
For repulsive interactions, the increase of SO coupling results in an
uncompressed less dense condensate and with increased multi-peak nature
of the density. 
The increase of Rabi coupling leads to a density increase with an almost 
constant number of multi-peaks.
For both signs of the interaction and negative values of Rabi coupling,
the condensate develops a notch in the central point and it seems to a 
dark-in-bright soliton.
In the case of the attractive nonlinearity, an interesting result is the
increase of the collapse threshold under the action of the SO and Rabi 
couplings.

\end{abstract}


\maketitle
 
\section{Introduction}

In last few years, the creation of synthetic non-Abelian gauge fields in
neutral atoms has attracted a great interest. 
Particularly, Lin \textit{et al.} \cite{SOC-1} accomplished the
experimental realization of an artificial spin-orbit (SO) coupling in a 
neutral atomic Bose-Einstein condensate (BEC) by means of 
counterpropagating laser beams coupling two atomic spin states of
$^{87}$Rb hyperfine state 5$S_{1/2}$, 
$|\uparrow \rangle  = |F=1 ,m_{F}=0 \rangle $ 
and $|\downarrow \rangle = |F=1 ,m_{F}=-1 \rangle $,
where $F$ is the total angular momentum of the hyperfine state and $m_F$ 
is the $z$ projection of $F$. In analogy with the two spin components of
a spin-half particle these states are called pseudo-spin-up and
pseudo-spin-down states.
This technique also has been implemented in Bose  \cite{soc-1} and atomic
Fermi gases \cite{SOC-Fermi}.
These experimental breakthroughs have led to significant theoretical 
works, opening the door to a fascinating and fast development of
unconventional phenomena on SO and Rabi coupled ultracold atoms research.
In \cite{Perspectives-SOC} are presented remarkable recent and
ongoing realizations of Rashba physics in cold atom systems, Dirac 
materials, Majorana femions and beyond.
Some basic aspects about quantum gases with SO coupling are introduced in
\cite{spin-orbit}.

The SO and Rabi coupled BECs have been studied in different contexts, 
solitons in one-dimensional (1D) geometries \cite{Solitons} and 
two-dimensional (2D) geometries \cite{2D-1}, vortices with or without
rotations \cite{Vortices}.
The quantum tricriticality and the phase transitions are studied by
considering a SO coupled configuration of spin-1/2 interacting bosons with
equal Rashba and Dresselhaus couplings \cite{Tricriticality}.
In \cite{order-disorder} it is shown that the bosons condenses into a
single momentum state of the Rashba spectrum via the mechanism of order 
by disorder.
Stability of plane waves in a 2D SO coupled BEC was studied analytically
in \cite{stability-1}.
A study of localization of a noninteracting and weakly interacting SO 
coupled BEC in a quasiperiodic bichromatic optical lattice (OL) potential
it was carried out in \cite{Adhikari-1}.
The existence of antiferromagnetically ordered (striped) ground states in 
a 1D SO coupled system with repulsive atomic interactions under the action
of a harmonic oscillator (HO) potential it is demonstrated in 
\cite{Malomed-1}.
The Josephson dynamics of a SO coupled BEC is investigated in
\cite{Josephson}.
There have also been efforts toward understanding SO coupling related 
physics at finite temperature \cite{T-finite}. 
Further, in trapped systems more intriguing ground state properties arise 
from the nontrivial interplay between SO coupling, confinement and
inter-atomic interactions.
The phase diagram includes two rotationally symmetric phases, a
hexagonally symmetric phase with triangular lattice of density minima
similar to that observed in rotating condensates, and a stripe phase
\cite{Santos-1}.
Wang \textit{et al.} have identified two different phases for the ground
state of homogeneous spin-1/2 and spin-1 BECs with Rashba SO coupling  
\cite{Spinor-BEC}.
From the spin-dependent interactions, the condensate is found to be a 
single plane wave or it is a standing wave forming spin stripes.
At strong SO coupling in trapped weakly interacting BECs, the single 
particle spectrum decomposes into discrete manifolds and quantum states
with Skyrmion lattices emerge when all bosons occupy the lowest manifold
\cite{vortex-5}. 
In ultracold Bose gases with Rashba SO coupling the presence of a weak 
harmonic potential results in a striped state with lower energy than the
conventional striped state in the homogeneous scenario 
\cite{Striped_states}.
The ground states of a single particle, two bosons, and two fermions
confined in a 1D harmonic trap with Raman-induced SO coupling are
considered in \cite{Harmon_trap}.
The collective dynamics of the SO coupled BEC trapped in a quasi-1D
harmonic potential it is studied in \cite{Collective_SOC}.
Recently, the influence of interaction parameters, SO and Rabi couplings
on the modulation instability in a condensate it is investigated in
\cite{MI-SOC}.
The combined effects of SO coupling and a non-local dipolar interaction
have given rise to a rich variety of ground state spin structures in a 
condensate \cite{Dipolar-SOC}.

At zero temperature, the time-dependent mean-field 3D Gross-Pitaevskii
equation (GPE) is a good theoretical tool for the study of dilute BECs
\cite{Dalfovo}. However, an interesting question is the derivation of
reduced 1D and 2D models to have an insight about the behavior of the 3D
full system.
Experimentally these effective models are useful for testing many-body
phenomena \cite{Bright-solitons1}. 
Theoretically they have been derived 1D and 2D models from the 3D GPE in
different contexts of BEC research \cite{1D-2D GPE}.
A starting point for the theoretical study of SO coupled BECs with reduced
dimensionality is provided by \cite{Salasnich_SOC1}, where a binary 
mean-field nonpolynomial Schr\"odinger equation (NPSE) has been used in
research of localized modes in condensates.
In a similar way, condensates with the SO coupling of mixed 
Rashba-Dresselhaus type and Rabi term are described by two coupled 2D
NPSEs \cite{Salasnich_SOC2}.

In this work we show the effect of harmonic traps on SO and
Rabi coupled BECs.
Starting from the 3D many-body Hamiltonian describing $N$ interacting 
bosons of equal mass with SO and Rabi couplings, we study the 1D reduction
of a 3D bosonic quantum field theory.
We achieve to getting two time-dependent 1D coupled nonpolynomial 
Heisenberg equations (NPHEs). 
If the many-body quantum state of the system is well-approximated to the
Glauber coherent state, our findings are consistent with those obtained
in a mean-field approximation \cite{Salasnich_SOC1}.
This is followed by a numerical solution of the coupled NPSEs.
For attractive and repulsive SO and Rabi coupled BECs we investigate some 
aspects of the interplay between the strength of the harmonic confinement,
the SO and Rabi couplings.
Our results show that the harmonic trap causes a reduction of the 
multi-peak nature of the condensate and it increases the density in both
attractive and repulsive condensates.
In the case of repulsive interactions, the increase of SO coupling results
in an increased multi-peak nature of the density with a less dense
condensate without its compression.
The increase of Rabi coupling leads to a density increase with an almost 
constant number of multi-peaks.
For both signs of the interaction and negative values of $\Gamma$, 
the condensate develops a notch in the central point and it seems to 
a dark-in-bright soliton \cite{dark-bright-soliton}.
A relevant finding is the increment of the strength of
the attractive nonlinearity at the collapse threshold.

The rest of the paper is organized in the following way.
In the Sec. \ref{II}, we derive two effective 1D coupled NPHEs and the
condition under which these match the 1D NPSEs describing BECs with SO and
Rabi couplings.
This is followed by a discussion about some aspects related to the 
numerical normalization of the wave-function in Sec. \ref{III}.
Numerical results of the SO and Rabi coupled BECs in harmonic traps are
reported in Sec. \ref{IV}. 
Finally, we present a summary and discussion of our study in Sec. \ref{V}.

\section{Theoretical model}
\label{II}

\subsection{Coupled Nonpolynomial Heisenberg Equations}
A quantum treatment of a dilute gas of $N$ interacting bosonic atoms of 
equal mass $m$, with SO and Rabi couplings can be obtained from the 
dimensionless many-body Hamiltonian given in \cite{SOC-Emerson}.
In order to obtain this dimensionless form, it has been used the HO 
length of the transverse trap $a_{\perp}=\sqrt{\hbar/(m\omega_{\perp})}$,
with the trapping frequency $\omega_{\perp}$.
The time $t$ is measured in units of $\omega^{-1}_{\perp}$, the spatial
variable ${\bf r}$ in units of $a_{\perp}$, the energy in units 
$\hbar \omega_{\perp}$ and the field operators are given in units of 
$a_{\perp}^{3/2}$. Such a Hamiltonian can be read as
\begin{eqnarray} 
{\hat H} &=& \int d{\bf r}
\sum_{j=1,2} \bigg \{ {\hat \psi}_{j}^\dagger
\bigg [ -{\frac{1}{2}}\nabla^{2}
+ U(\mathbf{r}) + (-1)^{j-1}i \gamma \frac{\partial}{\partial x} \bigg]
{\hat \psi}_{j}
\nonumber \\
&+& \Gamma {\hat \psi}_{j}^\dagger {\hat \psi}_{3-j} 
+ \sum_{l=1,2} \pi g_{jl} {\hat \psi}_{j}^\dagger{\hat \psi}_{l}^\dagger
{\hat \psi}_{j} {\hat \psi}_{l} \bigg \}
\label{H-field}
\end{eqnarray}
where ${\hat \psi_{j}}^\dagger$ and ${\hat \psi_{j}}$ $(j=1,2)$ are the
two pseudo-spin creation and annihilation boson field operators,
respectively. $g_{jj}\equiv 2a_{jj}/a_{\perp}$, and
$g_{12}=g_{21}\equiv 2a_{12}/a_{\perp}$ are the strengths of the intra-
and inter-species interactions, with $a_{jj}$ and $a_{12}$ the respective
intra- and inter-species s-wave scattering lengths.
$U\left(\mathbf{r}\right)$ is an external potential, and it is given as 
$U(\mathbf{r})= V(x) + (y^2 + z^2)/2$,
where $V(x)$ is a generic potential in the $x$ axial direction and the HO
potential keeps the confinement of the system in the transverse $(y, z)$
plane. $\gamma \equiv k_{L}a_{\perp}$ and 
$\Gamma  \equiv  \Omega /(2\omega_{\perp})$ are the dimensionless
strengths of the SO and Rabi couplings, respectively. $k_{L}$ is the 
recoil wave number induced by the interaction with the laser beams and
$\Omega$ is the frequency of the Raman coupling, responsible for the Rabi
mixing between the two states.
The bosonic field operators and their adjoints must satisfy the following
equal-time commutation rules, 
$ [ {\hat \psi}_{\alpha}({\bf r},t) , 
{\hat \psi}_{\beta}^{\dagger}({\bf r}',t) ] = \delta_{\alpha \beta}
\delta({\bf r}-{\bf r}')
\label{commuta-Bose-1} $
and
$ [ {\hat \psi}_{\alpha}({\bf r},t) ,
{\hat \psi}_{\beta}({\bf r}',t) ] = 
[ {\hat \psi}_{\alpha}^{\dagger}({\bf r},t) , 
{\hat \psi}_{\beta}^{\dagger}({\bf r}',t) ] = 0
\label{commuta-Bose-2} $,
where $\alpha, \beta =1,2$.
The creation of a particle in the state $|{\bf r},\alpha,t\rangle$ from 
the vacuum state $|0\rangle$ is given as 
${\hat \psi}_{\alpha}^{\dagger}({\bf r},t) 
|0\rangle = |{\bf r},\alpha,t\rangle$.
The annihilation of a particle in the state $|{\bf r},\beta,t\rangle$ is
given by
${\hat \psi}_{\alpha}({\bf r}',t) |{\bf r},\beta, t\rangle = 
\delta_{\alpha\beta} \delta({\bf r}-{\bf r}') |0\rangle$. 
From the Heisenberg equation of motion 
$ i {\partial_t {\hat \psi}_{j} }
= [ {\hat \psi}_{j} , {\hat H} ] 
\label{Heisenberg-eq}$, 
we get two coupled field equations in a closed form,
\begin{eqnarray}
i{\frac{\partial}{\partial t}}{\hat \psi}_{j}(\mathbf{r},t) 
&=&
\bigg[ -{\frac{1}{2}}\nabla^{2} + V(x)
+ {1\over 2} \left( y^2 + z^2 \right)
\nonumber 
\\
&+& (-1)^{j-1}i \gamma \frac{\partial}{\partial x}
+2\pi g_{jj} {\hat \psi}_{j}^{\dagger}(\mathbf{r},t)
{\hat \psi}_{j}(\mathbf{r},t)
\nonumber 
\\
&+& 2\pi g_{12} {\hat \psi}_{3-j}^{\dagger}(\mathbf{r},t)
{\hat \psi}_{3-j}(\mathbf{r},t)\bigg] {\hat \psi}_{j}
(\mathbf{r},t) 
+\Gamma {\hat \psi}_{3-j}(\mathbf{r},t)
\label{field-eq}
\end{eqnarray}

\subsection{1D reduction of the 3D Hamiltonian}
To perform the 1D reduction of the 3D Hamiltonian (\ref{H-field}), we 
suppose that the single-particle ground state in the transverse $(y, z)$
plane is given by a Gaussian wave-function 
\cite{SOC-Emerson, Salasnich-Quantum-solitons, Salasnich-QFT, 
Salasnich-book}, such that
\begin{equation}
{\hat \psi}_{j}({\bf r}) |G\rangle = 
{1\over \sqrt\pi \ \eta_{j}(x,t)}
\exp{\left[ - {y^2+z^2\over 2\eta^2_{j}(x,t)} 
\right] }\, {\hat \phi}_{j}(x,t) |G\rangle \; 
\label{supposing-1}
\end{equation}
where $|G\rangle$ is the many-body ground state, while 
${\hat \phi}_{j}(x,t)$
and $\eta_{j}(x,t)$ $(j=1,2)$ are, respectively, the axial bosonic field 
operators and the transverse widths.
Inserting this ansatz into the Hamiltonian (\ref{H-field}), performing the
integration in the transverse plane, and neglecting derivatives of 
$\eta_{j}(x,t)$, one can derive the corresponding effective 1D-Hamiltonian
${\hat h}_{1D}$ \cite{SOC-Emerson}, such that 
${\hat H}|G\rangle = {\hat h}_{1D} |G\rangle$.
The tranverse widths $\eta_{j}(x,t)$ are determined by minimizing the
energy functional $\langle G|{\hat h_{1D}}|G\rangle$ with respect to
$\eta_{j}(x,t)$ \cite{ SOC-Emerson,Salasnich-QFT}.
The ground state $|G\rangle$ is obtained self-consistently from
${\hat h}_{1D}$ and $\eta_{j}(x,t)$.
Now, from ${\hat h}_{1D}$ and 
$ i {\partial_{t} \hat \phi}_{j} = [ {\hat \phi}_{j} , {\hat h}_{1D} ]$,
we derive two 1D coupled NPHEs describing a many-body quantum system of
dilute bosonic atoms with SO and Rabi couplings,
\begin{eqnarray} 
i {\partial \over \partial t} {\hat \phi}_{j} &=&
\bigg [ -{\frac{1}{2}} \frac{\partial^2}{\partial x^2}
+ V(x) + \frac{1}{2} \left( \frac{1}{\eta_{j}^2} + 
\eta_{j}^2 \right)
+ (-1)^{j-1}i \gamma \frac{\partial}{\partial x} 
\nonumber \\
&+& \frac{g_{jj}}{\eta^{2}_{j}} 
{\hat \phi}_{j}^\dagger {\hat \phi}_{j} 
+ \frac{2 g_{12}}{\eta^2_{1} + \eta^2_{2}}
{\hat \phi}_{3-j}^\dagger {\hat \phi}_{3-j} \bigg]
{\hat \phi}_{j}
+ 2\Gamma \frac{\eta_{1} \eta_{2}}{\eta_{1}^2 + \eta_{2}^2}
{\hat \phi}_{3-j}
\label{1DNPSE-field}
\end{eqnarray}
This system of equations must be solved self-consistently using the
many-body quantum state of the system $|S\rangle$ in the equations of the
transverse widths $\eta_{j}$ \cite{ SOC-Emerson,Salasnich-QFT}. So
\begin{eqnarray}
\eta_{j}^4 &=&
1+ g_{jj} \frac{\langle S| {\hat \phi}_{j}^\dagger 
{\hat \phi}_{j}^\dagger
{\hat \phi}_{j} {\hat \phi}_{j} |S\rangle } 
{\langle S|
{\hat \phi}_{j}^\dagger {\hat \phi}_{j} |S\rangle}
+ 4g_{12} \frac{\eta_{j}^{4}}{(\eta^2_{1} + \eta^2_{2})^2}
\frac{\langle S| {\hat \phi}_{1}^\dagger {\hat \phi}_{2}^\dagger
{\hat \phi}_{2} {\hat \phi}_{1} |S\rangle } 
{\langle S|
{\hat \phi}_{j}^\dagger {\hat \phi}_{j} |S\rangle}
\nonumber \\
&+& 2(-1)^{j-1} \Gamma \eta_{j}^{3} \eta_{3-j} 
\frac{\eta_{1}^2 - \eta_{2}^2}{(\eta_{1}^2 + \eta_{2}^2)^2}
\frac {\langle S|  ({\hat \phi}_{1}^\dagger {\hat \phi}_{2} + 
{\hat \phi}_{2}^\dagger {\hat \phi}_{1} )|S\rangle}
{\langle S|{\hat \phi}_{j}^\dagger {\hat \phi}_{j}|S\rangle}
\label{widths-field}
\end{eqnarray}
As a particular case with $\eta_{j}^4= 1$, the model (\ref{1DNPSE-field})
describes an one-dimensional dilute gas of bosonic atoms with SO and Rabi 
couplings confined by a generic potential $V(x)$ in the $x$ direction.

Nevertheless, in an open environment particles may be added or removed 
from the system and one can suppose that the number of bosons in the
matter field is not fixed, i.e. the system is not in a pure Fock state
\cite{Salasnich-book,Coherent-states-1,Coherent-states-QFT}.
Then in analogy with a functional-integral formalism to investigate
interacting quantum gases \cite{Coherent-states-QFT} and the description 
in optics of a radiation field of a laser device operating in a single 
mode \cite{Coherent-states-2}, we introduce the Glauber coherent state
$|GCS\rangle$, which does not have a fixed number of particles.
This coherent state is defined as an eigenstate of the annihilation 
operator. In terms of quantum field operators we have
${\hat \phi}_{j}(x,t)|GCS\rangle = \phi_{j}(x,t)|GCS\rangle$ where
$\phi_{j}(x,t)$ is a classical field, and the average number of atoms in
the coherent state is given as $N_{j}=\langle GCS|\hat N_{j}|GCS\rangle$.
Hence, when the many-body quantum state $|S\rangle$ is considered as a
Glauber coherent state $|GCS\rangle$ 
\cite{Salasnich-Quantum-solitons, Salasnich-QFT, Salasnich-book},
the 1D NPHE (\ref{1DNPSE-field}) becomes 1D NPSE
\cite{Salasnich_SOC1, SOC-Emerson}
\begin{eqnarray} 
i {\partial \over \partial t} \phi_{j} &=&
\bigg [ -{\frac{1}{2}} \frac{\partial^2}{\partial x^2}
+ V(x) + \frac{1}{2} \left( \frac{1}{\eta_{j}^2} + 
\eta_{j}^2 \right)
+ (-1)^{j-1}i \gamma \frac{\partial}{\partial x}
\nonumber \\
&+& \frac{g_{jj}}
{\eta^{2}_{j}} 
\left|\phi_{j}\right|^2
+ \frac{2 g_{12}}{\eta^2_{1} + \eta^2_{2}}
\left|\phi_{3-j}\right|^2 \bigg]
\phi_{j}
+ 2\Gamma \frac{\eta_{1} \eta_{2}}{\eta_{1}^2 +
\eta_{2}^2} \phi_{3-j}
\label{NPSE-SOC}
\end{eqnarray}
The time-dependent spinor complex wave-functions $\phi_{j}$ $(j=1,2)$
describe the two pseudo-spin components $|\uparrow \rangle$ and 
$|\downarrow \rangle$, respectively.
The normalization conditions are given by
$\int_{-\infty}^{\infty}dx|\phi_{j}(x,t)|^2=N_{j}$,
where $N_{j}$ is the number of atoms in the  $j$ component, and the 
conserved total number of atoms is $N=N_{1}+N_{2}$.
The corresponding transverse widths (\ref{widths-field}), become
\begin{eqnarray}
\eta_{j}^4 &=&
1+ g_{jj} \left|\phi_{j}\right|^2
+ 4g_{12} \frac{\eta_{j}^{4}}{(\eta^2_{1} + \eta^2_{2})^2}
\left|\phi_{3-j}\right|^2
\nonumber \\
&+& 2(-1)^{j-1}\Gamma \eta_{j}^{3} \eta_{3-j} 
\frac{\eta_{1}^2 - \eta_{2}^2}{(\eta_{1}^2 + \eta_{2}^2)^2}
\frac { (\phi_{1}^{*} \phi_{2} + \phi_{2}^{*} \phi_{1})}
{\left|\phi_{j}\right|^2 }
\label{widths}
\end{eqnarray}
In general, the coupled Eqs. (\ref{NPSE-SOC}) are strictly 1D under the 
condition $\eta_{j}^4= 1$ in the system of Eqs. (\ref{widths}), 
establishing the conventional 1D GPE with SO and Rabi couplings
\begin{eqnarray} 
i {\partial \over \partial t} \phi_{j} &=&
\bigg [ -{\frac{1}{2}} \frac{\partial^2}{\partial x^2}
+ V(x) + (-1)^{j-1}i \gamma \frac{\partial}{\partial x} 
\nonumber \\
&+& g_{jj} \left|\phi_{j}\right|^2
+ g_{12} \left|\phi_{3-j}\right|^2 \bigg] \phi_{j} 
+\Gamma \phi_{3-j}
\label{1D-eq}
\end{eqnarray}
we have omitted the constant contribution of the transverse energy given 
as $1$ (in units of $\omega_{\perp}$).
This model becomes the one that it was implemented to studying nonlinear
modes in binary bosonic condensates with nonlinear repulsive interactions
and linear SO- and Zeeman-splitting couplings \cite{Malomed-1}.
For this purpose, it is necessary to use the transformations
$(-1)^{j-1} \partial_{x} \phi_{j} \rightarrow \partial_{x} \phi_{3-j}$
and $\phi_{3-j} \rightarrow (-1)^{j+1}\phi_{j}$, which are tantamount to 
the global pseudo-spin rotations $\sigma_{x}\rightarrow \sigma_{z}$ and
$\sigma_{z}\rightarrow \sigma_{x}$ given in \cite{SOC-Emerson}.
Another useful conserved quantity relating the two components of the 
density is the pseudo-magnetization density $\mathcal{M}$ defined as
$\mathcal{M} \equiv N^{-1} \left(|\phi_{1}(x)|^2 -|\phi_{2}(x)|^2 \right)$ 
\cite{Malomed-1,normalization}, with the respective total 
pseudo-magnetization $M=\int_{-\infty}^{\infty} dx \mathcal{M}$.
The conserved total number of atoms and the conserved total
pseudo-magnetization are considered as constraints to compute the ground 
state of SO and Rabi coupled BECs as it is discussed in Sec. \ref{III}.

The above models (\ref{NPSE-SOC}) and (\ref{widths}) were previously 
derived in a mean-field approximation  describing a BEC with pseudo-spin
states $|\uparrow \rangle$ and $|\downarrow \rangle$ 
\cite{Salasnich_SOC1}. 
In this approximation the number of atoms in the single-particle condensed
state is considered large and the quantum fluctuations are negligible so
that the Bogoliubov approximation can be used \cite{Dalfovo}. 
Indeed our results let see the correspondence between quantum field theory 
and classical field theory. Furthermore, these allow suggest that the
models obtained using the Glauber coherent state are tantamount to the
Bogoliubov approximation in the study of BECs with SO and Rabi couplings
at zero temperature.

In the full symmetric case, i.e. $g_{11}=g_{22}=g_{12}\equiv g$, 
Eqs. (\ref{widths}) take the form
$\eta_{1}^4=\eta_{2}^4= 1+ g(\left|\phi_{1}\right|^2 + 
\left|\phi_{2}\right|^2)$, where it is worth noting that $g<0$ and $g>0$
describe self-attractive and self-repulsive binary BECs, respectively.
In that way, the system of two coupled equations (\ref{NPSE-SOC}) can be
read as \cite{Salasnich_SOC1, SOC-Emerson}
\begin{eqnarray} 
i {\partial \over \partial t} \phi_{j} &=&
\Bigg [ -{\frac{1}{2}} \frac{\partial^2}{\partial x^2}
+ V(x) + (-1)^{j-1}i \gamma \frac{\partial}{\partial x} 
\nonumber \\
&+& \frac{1 + (3/2)g(\left|\phi_{1}\right|^2 + 
\left|\phi_{2}\right|^2)}{\sqrt{1 + g(\left|\phi_{1}\right|^2 + 
\left|\phi_{2}\right|^2)}} \Bigg] \phi_{j} + \Gamma 
\phi_{3-j}
\label{NPSE-SO-Rabi}
\end{eqnarray}
From Eq. (\ref{NPSE-SO-Rabi}), we construct stationary states taking the 
chemical potential $\mu$, and setting 
$\phi_{j}(x,t) \rightarrow \sqrt{N} \phi_{j}(x)\exp{(-i\mu t)}$.
The resulting equations for stationary fields $\phi_{1}$ and $\phi_{2}$ 
are compatible with restriction
\begin{eqnarray} 
\phi_{1}^*(x)=\phi_{2}(x)
\label{restriction}
\end{eqnarray}
Now, we establish the transformation $\phi_{1}(x) = \Phi(x)/\sqrt{2}$, and
along with the above mentioned restriction between the stationary fields,
we get a single stationary NPSE with SO and Rabi couplings
\cite{Salasnich_SOC1,SOC-Emerson}
\begin{eqnarray} 
\mu \Phi &=&
\bigg [ -{\frac{1}{2}} \frac{d^2}{d x^2}
+ V(x) + i \gamma \frac{d}{d x} 
+ \frac{1 + (3/2)gN |\Phi|^2}{\sqrt{1 + gN |\Phi|^2}}
\bigg] \Phi + \Gamma \Phi^{*}
\label{Stationary-NPSE}
\end{eqnarray}
so that the normalization is $\int_{-\infty}^{\infty} dx |\Phi(x)|^2=1$.
Henceforth, we use the harmonic axial potential $V(x)=\lambda^{2}x^{2}/2$,
with frequency $\omega_x$ and anisotropy 
$\lambda\equiv\omega_{x}/\omega_{\perp}$.
We note that, in the strong nonlinear regime $gN|\Phi|^2 \gg 1$, 
the nonpolynomial nonlinearity reduces to a quadratic form
$(3/2)\sqrt{gN}|\Phi|\Phi$.
On the other hand, in the weakly nonlinear regime $gN|\Phi|^2 \ll 1$,
the nonlinearity takes the cubic form $ gN|\Phi|^2\Phi$. 
In the two approaches without loss of generality we have omitted the
transverse contribution.
In absence of SO coupling $\gamma=0$, the solutions of Eq. 
(\ref{Stationary-NPSE}) are real and the resulting equation is tantamount
to the usual version of the NPSE with a shifted chemical potential, 
$\mu \rightarrow \mu-\Gamma$.
In general, solutions of Eq. (\ref{Stationary-NPSE}) are complex if
$\gamma \neq 0$.

\section{The normalization constants}
\label{III}

In the presence of different harmonic traps, we study the ground state
of SO and Rabi coupled BECs by numerically solving the stationary 1D NPSE
(\ref{Stationary-NPSE}).
In order to find the ground state, we use a split-step Crank-Nicolson 
method with imaginary time propagation \cite{Adhikari-numeric}.
In the imaginary time propagation $(t\rightarrow-it)$, the time evolution 
operator is not unitary and we have not neither conservation of the
normalization nor the magnetization.
In \cite{Salasnich_SOC1, SOC-Emerson} the conserved total number of atoms
is considered as a unique constraint to numerically find the ground state
of the condensate and the conserved total magnetization was not taken into
account. In this paper, using the conservation of these two quantities we 
develop a way to calculate the normalization constants.
To fix both the normalization and the magnetization we propose the
following approach to renormalize the wave-function after each operation 
of Crank-Nicolson method. 
We consider the continuous normalized gradient flow discussed in 
\cite{normalization} and used in the study of SO coupled spin-1
\cite{Adhikari-2} and spin-2 \cite{Adhikari-3} BECs.
So after each iteration the wave-function components in Eq.
(\ref{NPSE-SOC}) are transformed as
$\phi_{j}(x,t + \Delta t) = d_{j}\phi_{j}(x,t)$
$(j=1,2)$, where $d_{j}$ are the normalization constants.
Now, the constraint on the total number of atoms can be written in terms
of $d_{j}$'s and the wave-function components $N_{j}$ as 
$\sum_{j=1,2} d_{j}^{2}N_{j}=N$, where 
$N_{j}= \int_{-\infty}^{\infty} dx |\phi_{j}(x)|^{2}$.
Here we have two unknowns $d_{1}$ and $d_{2}$, and only one equation given
by the condition on $N$.
In order to determine the values of the normalization constants $d_{j}$
using the definition of the pseudo-magnetization density provided in 
Sec. \ref{II} we introduce the constraint on the total
pseudo-magnetization as
$N^{-1} \sum_{j=1,2} (-1)^{j-1}d_{j}^{2}N_{j} = M$.
Without loss of generality on $N$ or $M$, we set the variable change 
$\phi_{j} \rightarrow  \sqrt{N}\phi_{j}$, and we obtain the nonlinear
system of equations,
\begin{eqnarray} 
\sum_{j=1,2} d_{j}^{2} \int_{-\infty}^{\infty} dx |\phi_{j}(x)|^{2} = 1
\label{normaliz1}
\end{eqnarray}
\begin{eqnarray} 
\sum_{j=1,2} (-1)^{j-1}d_{j}^{2}\int_{-\infty}^{\infty} dx
|\phi_{j}(x)|^{2} = M
\label{normaliz2}
\end{eqnarray}
Solving this system of equations, we get explicitly the normalization 
constants
\begin{eqnarray} 
d_{j}= \sqrt{\frac{1 + (-1)^{j-1}M}{2\int_{-\infty}^{\infty} dx
|\phi_{j}(x)|^{2} }}
\label{coeffs}
\end{eqnarray}
Particularly, we consider the stationary solutions of the coupled Eqs. 
(\ref{NPSE-SO-Rabi}) by numerically solving the Eq. (\ref{Stationary-NPSE}). 
As result of restriction (\ref{restriction}), the total 
pseudo-magnetization $M$ is zero, and the required normalization constants
(\ref{coeffs}) take the form
\begin{eqnarray} 
d_{1}=d_{2}\equiv d =
\frac{1}{\sqrt{2\int_{-\infty}^{\infty} dx |\Phi(x)|^2}}
\end{eqnarray}

\section{Numerical results}
\label{IV}

We split the stationary 1D NPSE (\ref{Stationary-NPSE}) such that, the 
kinetic term, the SO contribution, the HO potential and the nonpolynomial
nonlinearity are discretized and implemented as it is presented in
\cite{Adhikari-numeric}.
To ensure the conservation of both the total number of atoms and the 
magnetization after each operation of Crank- Nicolson method, we use the
new normalization constants $d_{j}$ found in the previous section.
By adjusting the numerical method used to obtaining the ground state of
atomic-molecular BECs \cite{molecular-BEC-numeric}, the Rabi contribution 
is handled.
The spatial and time steps employed in the present work are
$\Delta x=7.31\times 10^{-3}$ and $\Delta t=1.51\times 10^{-4}$
respectively.
As initial input in the numerical simulations we consider the normalized
Gaussian wave-fuction
$\Phi=(\pi \omega_x ^2)^{-1/4}\exp(-x^{2}/2\omega_x ^2)$,
where $\omega_x$ is the width along $x$ axis.

\begin{figure}[b] 
\begin{center}
\includegraphics[width=11cm,clip]{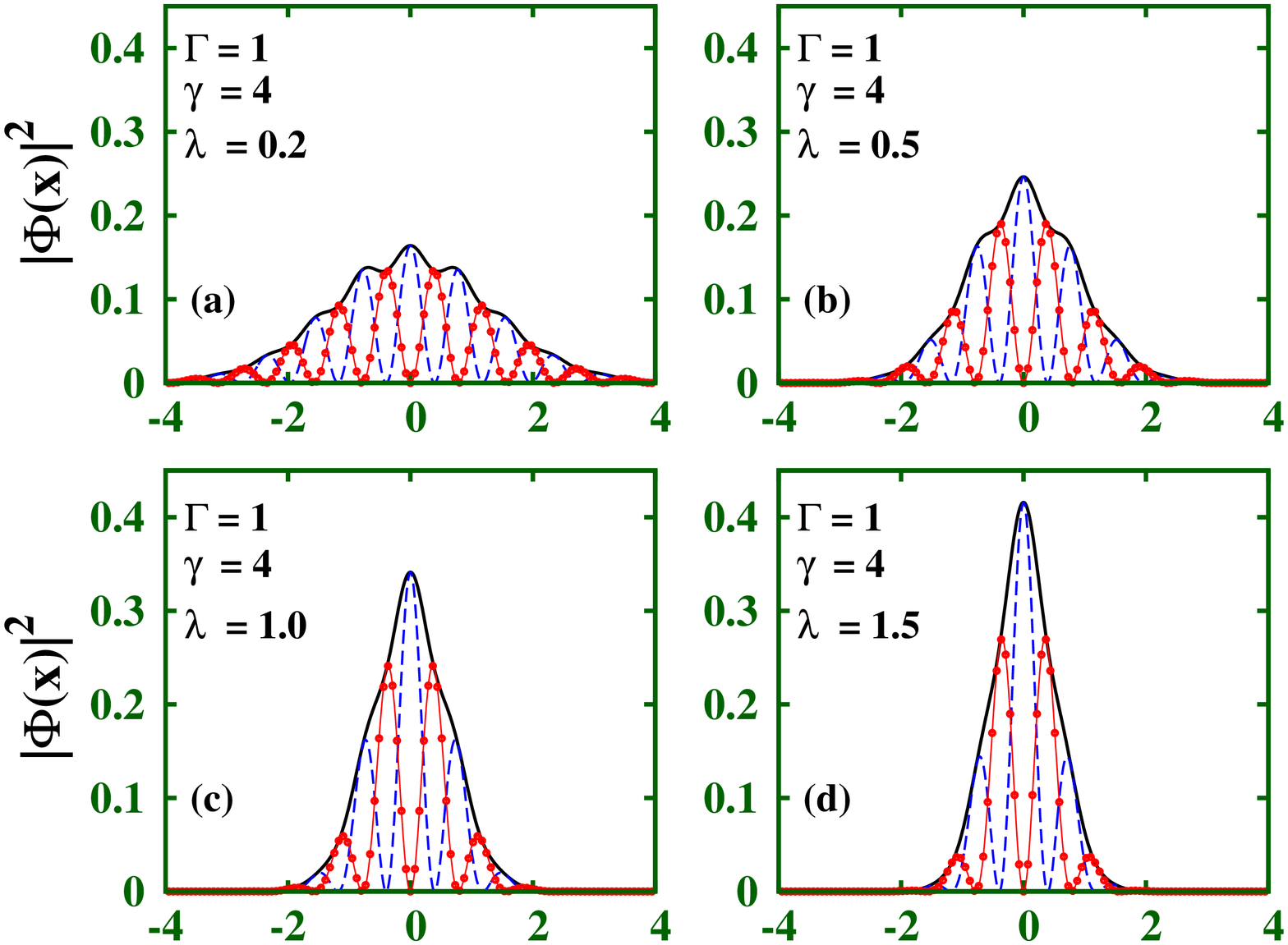}
\includegraphics[width=11cm,clip]{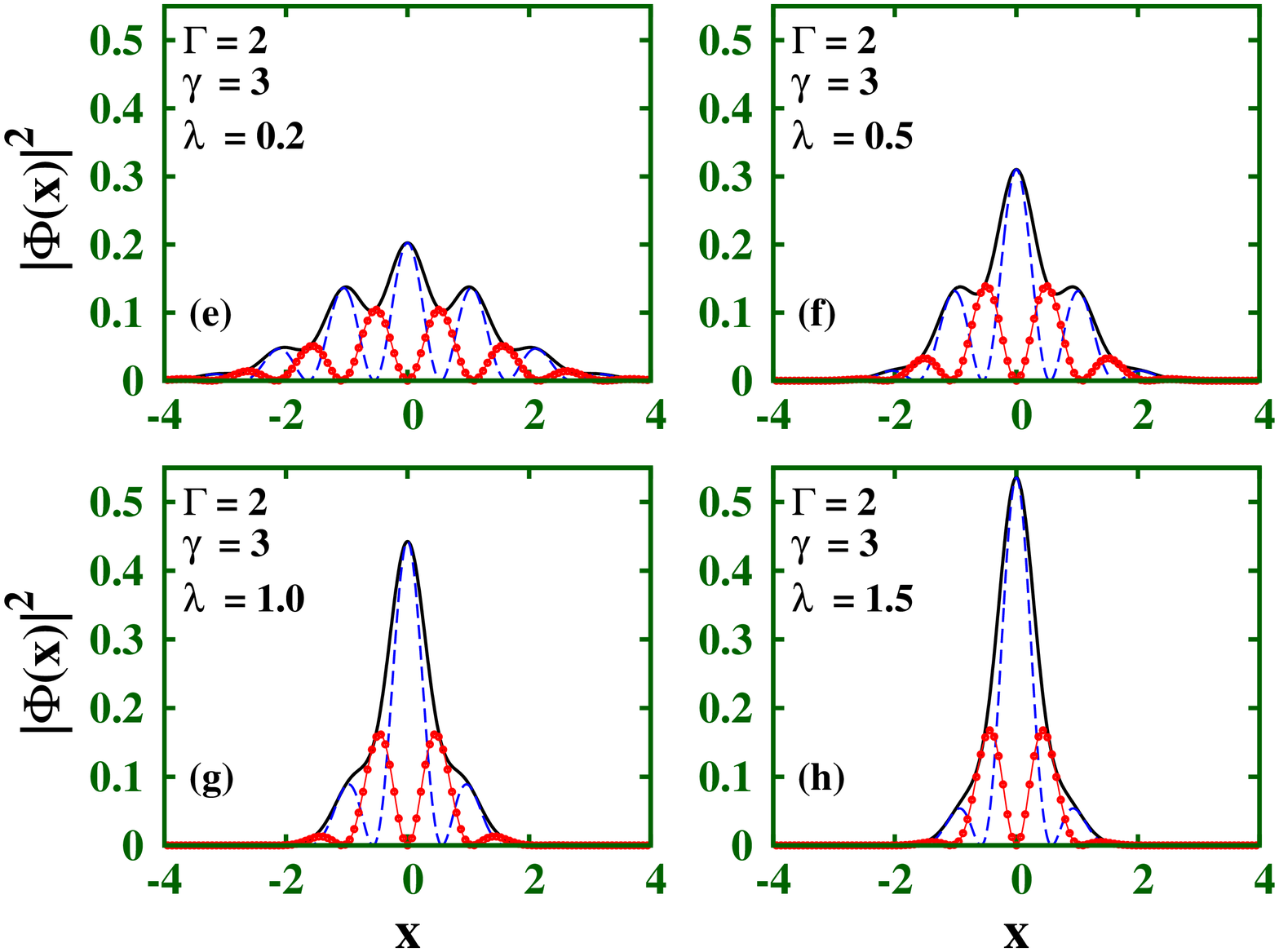}
\end{center}
\caption{(Color online)
Density profile of a self-attractive SO and Rabi coupled BEC under 
axial harmonic confinement $V(x)=\lambda^{2}x^{2}/2$. The nonlinearity 
strength is $gN=-0.8$. For panels (a)-(d), we set the Rabi coupling 
$\Gamma=1$, the SO coupling $\gamma=4$,  and four values of the trap
strength $\lambda$. For panels (e)-(h) we have $\Gamma=2$, $\gamma=3$, 
and the same values of $\lambda$ as in panels (a)-(d). The solid line
depicts the density, while the dashed and point lines represent
squared real and imaginary parts of the wave-function, respectively. Lengths are
measured in units of the transverse confinement radius
$a_{\perp}=\sqrt{\hbar/(m\omega_{\perp})}$.}
\label{Attractive1}
\end{figure}

\begin{figure}[t] 
\begin{center}
\includegraphics[width=11cm,clip]{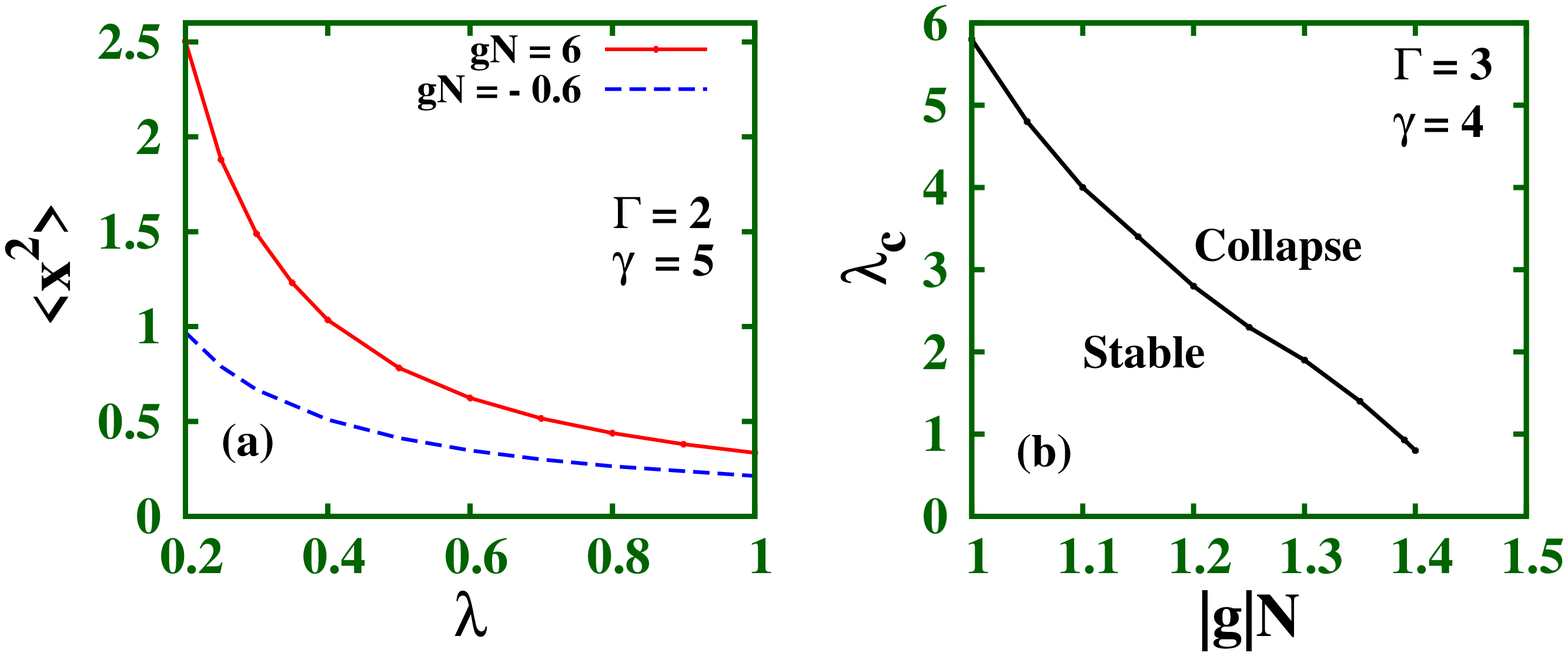}
\end{center}
\caption{(Color online) 
(a) Root mean square (rms) size of the condensate $\langle x^2 \rangle$
versus the trap strength $\lambda$. Here $\Gamma=2$, $\gamma=5$, and two
values of the nonlinearity strength $gN$.
(b) The critical trap strength $\lambda_{c}$ versus the
attractive nonlinearity strength $|g|N$ with $\Gamma=3$ and $\gamma=4$.
Lengths are measured in units of $a_{\perp}$.}
\label{collapse}
\end{figure}

We start the numerical analysis of the ground state structure in
self-attractive BECs $(g<0)$ with SO and Rabi couplings in presence of a
HO axial potential $V(x)=(\lambda^{2}/2)x^2$.
In absence of the axial trapping potential $V(x)=0$, the existence of
bright solitons as solutions of Eq. (\ref{Stationary-NPSE}) is considered
in \cite{Salasnich_SOC1, SOC-Emerson}. 
In \cite{Salasnich_SOC1} is shown that for $\gamma$ and $-gN$ constants,
the increase of $\Gamma$ results in a compressed density profile with an
increase of its height.
On the other hand, by setting $\Gamma$ and $-gN$, the bright soliton is
not subject to a significant compression, its height is reduced and the
number of local maxima is increased with the increase of $\gamma$
\cite{SOC-Emerson}.
In Fig. \ref{Attractive1} we display the numerical results of a  
harmonically trapped self-attractive SO and Rabi coupled BEC.
The density profile $|\Phi(x)|^2$, is plotted as a function of axial
coordinate $x$.
In panels (a)-(d), we set the strength of the nonlinearity $gN=-0.8$, the
Rabi coupling $\Gamma=1$, the SO coupling $\gamma=4$, and four values of 
the strength of the trap $\lambda$. 
In panels (e)-(h), we set $\Gamma=2$, $\gamma=3$, and the same values of
$gN$ and $\lambda$ as in panels (a)-(d). 
With the increase of the trap strength, the modulations of the squared of 
the real part and the imaginary one are subjected to a compression and
their heights are increased.
A noteworthy feature is a further increase of real part in comparison 
with the increase of the imaginary one.
As a consequence of the effect caused by the trap strength the arrangement
of the atoms in the condensate is restricted, their oscillations and its 
multi-peak nature decrease, thus giving rise to a shrunken condensate with
a higher density.
Reducing $\gamma$ and increasing $\Gamma$ at the same time, panels
(a)-(e), (b)-(f), (c)-(g) and (d)-(h), our findings show the possibility 
of reduce the number of modulations and enlarge the density without a 
significant compression of the condensate.
We compare the results of 1D NPSE (\ref{Stationary-NPSE}) with those of 
the stationary 1D GPE obtained from (\ref{1D-eq}) and both models match 
very good each other. The coincidence between these models can be
understood since the 1D NPSE with $gN = -0.8$ becomes effectively 1D, i.e.
the 1D NPSE becomes 1D GPE. 
However it is worth noting that, contrary to the cubic nonlinearity of
1D GPE, the nonpolynomial term of 1D NPSE is quite essential to predict 
the instability by collapse. This interesting phenomenon is present as
long as the density of the condensate reaches the critical value
$|\Phi|^{2}=(|g|N)^{-1}$ \cite{Salasnich_SOC1, SOC-Emerson}.
The normalization constants obtained in Sec. \ref{III} let us to get
the same behaviors presented in \cite{Salasnich_SOC1, SOC-Emerson} but
with a lower density.
This density decrease gives rise to a greater variation of the parameters
in particular in attractive SO and Rabi coupled BECs where the presence of
instability by collapse is affected as shown below in the analysis 
of Fig. \ref{collapse} (b).

\begin{figure}[b] 
\begin{center}
\includegraphics[width=11cm,clip]{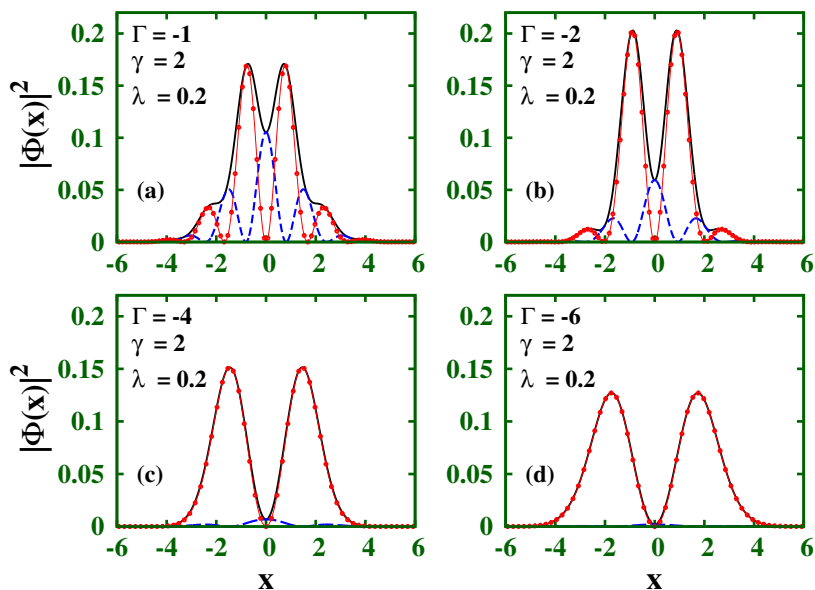}
\end{center}
\caption{(Color online)
Density profile of a self-attractive SO and Rabi coupled BEC under the
axial harmonic confinement $V(x)=\lambda^{2}x^{2}/2$.
Here, $gN=-0.8$, $\lambda=0.2$, $\gamma=2$ and four negative values of 
$\Gamma$.
The solid line depicts the density, while the dashed and point lines
represent squared real and imaginary parts of the wave-function, 
respectively. Lengths are measured in units of $a_{\perp}$.}
\label{Attractive2}
\end{figure}

Next we consider a harmonically trapped attractive condensate with a
negative value of Rabi coupling. In Fig. \ref{Attractive2} (a)-(d), we
plot the density profile with the same nonlinearity as in Fig.
\ref{Attractive1}, with a weak trap strength $\lambda=0.2$, the SO 
coupling $\gamma=2$ and four values of Rabi coupling $\Gamma$. 
Upon numerical simulation with the increase of $\Gamma(<0)$ the peaks of 
the density are diminished, the real component of the wave-function is
suppressed Fig. \ref{Attractive2} (d) and such a condensate is found to 
develop a notch in the central point $x=0$ between the two halves in a 
similar way as in the density distribution of a dark soliton.
In Fig. \ref{Attractive2} (b)-(c), the condensate seems to a grey-like 
soliton with the central notch having nonzero density.
Using the same nonlinearity as in Fig. \ref{Attractive1}, setting the 
values of $\gamma$ and $\Gamma(<0)$,
the increase of trap strength causes a compression of the condensate, its 
density becomes higher and the number of oscillations is reduced.
All of this in a similar way as it was described in 
Fig. \ref{Attractive1}.
About the symmetry of the system an interesting issue arises.
As is predicted in \cite{Salasnich_SOC1}, without the axial confinement 
and $\Gamma<0$, the real and imaginary parts of the wave-function are odd 
and even, respectively.
In presence of the axial trapping potential and regardless of the sign of
$\Gamma$ the real and imaginary parts of wave-function are even 
and odd, respectively.
\begin{figure}[t] 
\begin{center}
\includegraphics[width=7.8cm,clip]{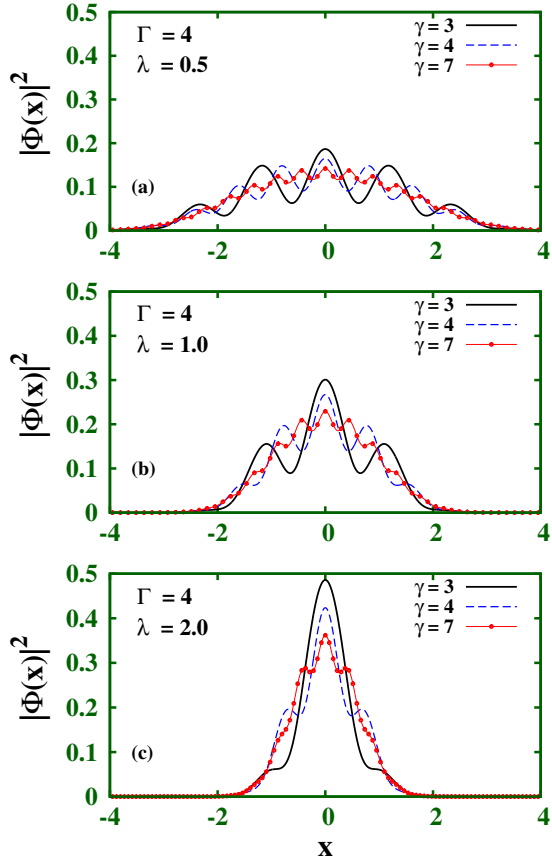}
\end{center}
\caption{(Color online)
Density profile of a harmonically trapped repulsive BEC with SO and Rabi 
couplings.
We use the same form of the confinement as in Fig. \ref{Attractive1},
$gN=12$, $\Gamma=4$, $\gamma=3,4,7$ and three different values of
$\lambda$, panels (a)-(c).
Lengths are measured in units of $a_{\perp}$.}
\label{Repulsive1}
\end{figure}
\begin{figure}[t] 
\begin{center}
\includegraphics[width=11cm,clip]{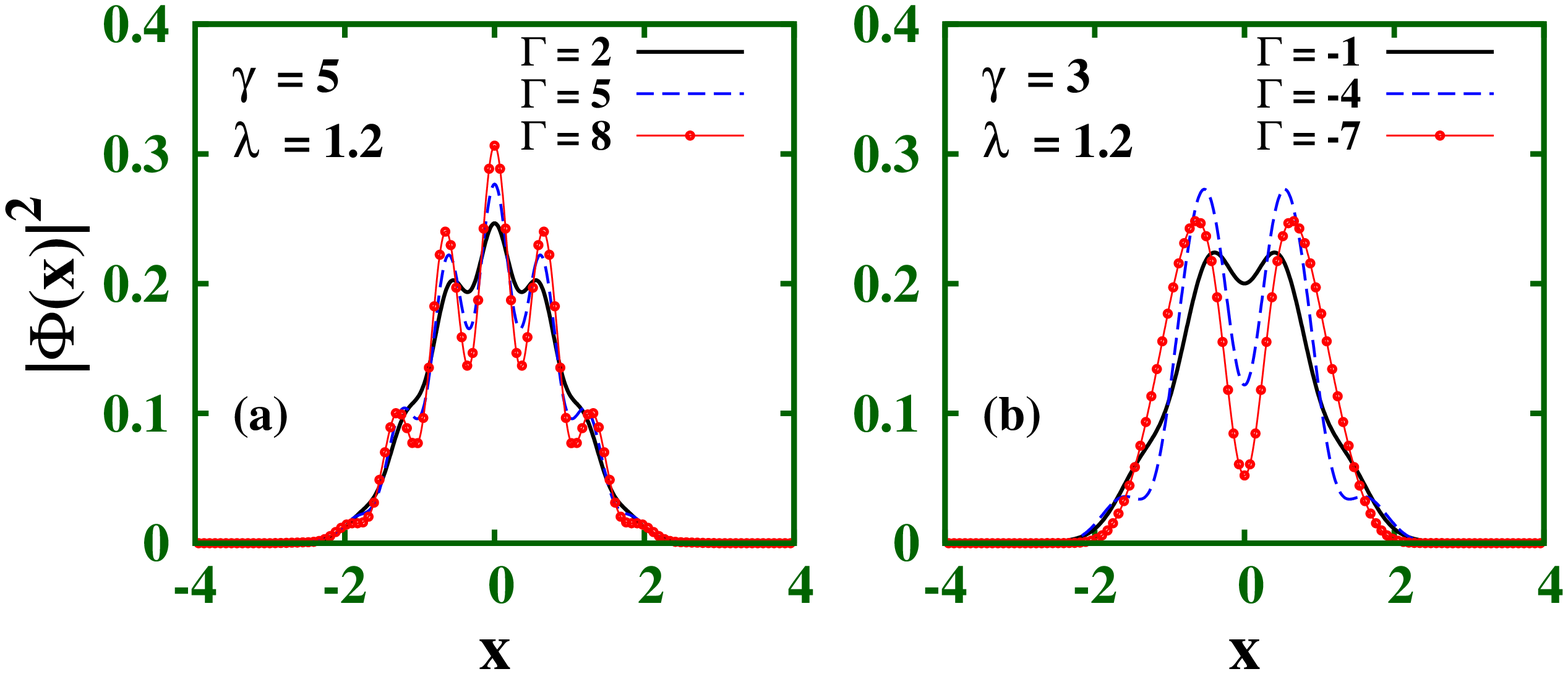}
\end{center}
\caption{(Color online) 
Density profile of a harmonically trapped repulsive BEC with SO and Rabi 
couplings.
We use $gN=16$ and the same form of the confinement as in Fig.
\ref{Attractive1} with $\lambda=1.2$. 
(a) $\gamma=5$ and $\Gamma=2,5,8$. (b) $\gamma=3$ and $\Gamma=-1,-4,-7$.
Lengths are measured in units of $a_{\perp}$.}
\label{Repulsive2}
\end{figure}

It is also relevant to analyze a self-repulsive $(g>0)$ BEC with SO and
Rabi couplings. If $V(x)=0$ the atoms are spread out and the condensation
is not possible. So a confinement potential is required to stabilize the
condensate. Here we use a harmonic potential $V(x)=(\lambda^{2}/2)x^2$.
In Fig. \ref{Repulsive1} (a)-(c), we plot the density $|\Phi(x)|^2$ as a
function of axial coordinate $x$, with $gN=12$, $\Gamma=4$, $\gamma=3,4,7$
and three different values of $\lambda$.
The increase of  $\lambda$ generates the same effect on the total density
as it was presented in Figs. \ref{Attractive1} and \ref{Attractive2}.
However for each of panels of Fig. \ref{Repulsive1}, with constant values
of $gN$, $\lambda$ and $\Gamma$, the interplay between the increase of SO 
coupling $\gamma$ and $\Gamma$ is reflected in two ways.
The increase of $\gamma$ provides a linkage between the atoms of the two
atomic hyperfine states, which in turn allows the increase of the 
multi-peak nature of the density. 
We also have a rearrangement of the atoms inside the trap, resulting in a 
less dense condensate and without its compression.

In Fig. \ref{Repulsive2} we plot the density of a trapped SO and Rabi
coupled BEC with repulsive nonlinearity $gN=16$ $(g>0)$
and the strength of trap $\lambda=1.2$.
In Fig. \ref{Repulsive2} (a), the increase of mixing between the two
states accounting for $\Gamma$ leads to a rearrangement of the atoms
without a significant change in the number of oscillations. 
While the real part increases the imaginary one undergoes a reduction
without the appearance of new peaks in none of both contributions.
So we have a density increase with an almost constant number of
multi-peaks. 
Due to increase of $\Gamma(<0)$ for a constant value of $\gamma$
Fig. \ref{Repulsive2} (b), the condensate develops a notch in the central
point and its seems to a dark-in-bright soliton with the central notch 
having nonzero density \cite{dark-bright-soliton}.
Our simulations show that the increase of $\lambda$ gives 
rise to a compressed and densest condensate as it has been presented in 
the above analysis of Figs. \ref{Attractive1} and \ref{Attractive2}.

The numerical results show that with the further increase of $\lambda$, 
the root mean square (rms) size of the condensate tends asymptotically to 
zero for both attractive $(g<0)$ and repulsive $(g>0)$ SO and Rabi coupled
BECs. This trend is plotted in Fig. \ref{collapse} (a).
In the range  $0 \leq\lambda < 0.2$ the attractive condensate remains.
For $\lambda=0$ we have a bright SO and Rabi coupled BEC soliton.
In the repulsive regime, the condensate is not present for $\lambda=0$,
and its rms diverges as $\lambda \rightarrow 0$.
In Fig. \ref{collapse} (b), we present a numerical stability diagram
for an attractive SO and Rabi coupled BEC with $\gamma=4$ and $\Gamma=3$.
For an attractive interatomic interaction the condensate is stable for
a critical maximum strength of interaction.
When the  strength of interaction increases beyond this critical value
the condensate becomes unstable and it collapses.
As usual, the experimental control of $g$ is achieved through
manipulation of the s-wave scattering length $a$ using a Feshbach
resonance.
Of our numerical findings beyond the threshold $|g|N = 1.4$ the 
instability by collapse occurs.
On the other hand, for a fixed interatomic interaction and the 
manipulation of trap strength the SO and Rabi coupled BEC is stable  
for $\lambda < \lambda_{c}$. Even the condensate is stable in the range 
$0 \leq \lambda < 0.8$ provided that $|g|N \leq 1.4$. 
These results express the relevance of the effective nonpolynomial 
nonlinearity of Eq. (\ref{Stationary-NPSE}) in the prediction of collapse.
In \cite{Salasnich_SOC1} at fixed values of SO coupling $(\gamma=1)$ and
Rabi coupling $(\Gamma=0.19)$, the collapse is predicted beyond 
$|g|N=1.2$.
Here we show the possibility of increasing the threshold value of the
collapse due to consideration of magnetization as a constraint 
in the calculation of the new normalization constants of Sec. \ref{III}.
So the present study could be useful to give an idea of the maximum number
of atoms in a stable SO and Rabi coupled BEC and could even be useful in 
planning future experiments.

The density patterns in this work for both attractive and repulsive 
BEC with SO and Rabi couplings in presence of harmonic traps are
symmetric with respect to the $\gamma<0$ scenario.
A possible verification of the results obtained in the present
work perhaps it could be carried out taking into account the recent
experimental realization of 2D SO coupling for BECs
\cite{SOC-2D-experimental} and the experimental adjustment of
$\Gamma$ and $\gamma$ carried out through of technique developed in
\cite{Tunable-SOC}. Here is proposed a scheme for controlling SO coupling 
between two hyperfine ground states in a binary BEC through a fast and
coherent modulation of the Raman laser intensities. Thus the experimental
manipulation of the Raman coupling tunes the SO coupling.

\section{Summary and outlook}
\label{V} 

In a bosonic quantum field theory we have derived an effective 1D system
of two coupled NPHEs describing a harmonically confined dilute gas of
bosonic atoms with nonlinear inter-atomic interactions, SO and Rabi 
couplings. The model takes into account both repulsive and attractive
inter-atomic interactions. 
Provided that the many-body quantum state of the system is assumed to 
well-approximated by the Glauber coherent state, the 1D coupled NPHE 
becomes 1D coupled NPSE.

We focus on numerical analysis of the 1D coupled NPSE describing 
harmonically trapped attractive and repulsive SO and Rabi coupled BECs.
The harmonic trap causes a strong reduction of the multi-peak nature of 
the condensate and it increases the density in both attractive and
repulsive SO and Rabi coupled BECs.
In the case of the repulsive interactions, the increase of $\gamma$
results in a less dense condensate without its compression and with a 
increase of the multi-peak nature of the density.
On the other hand, the increase of $\Gamma$ leads to a density increase
with an almost constant number of multi-peaks.
Numerically we found a new structure of the SO and Rabi coupled BECs for
both signs of the interaction and negative values of $\Gamma$.
Here the condensate develops a notch in the central point and it seems to 
a dark-in-bright soliton with the central notch having nonzero density
\cite{dark-bright-soliton}.
An interesting and significant result is the increment of the strength of
the attractive nonlinearity at the collapse threshold due to the use
of the magnetization as a constraint in the ground state calculation of 
the condensate under the combined action of the SO and Rabi couplings.

Extending the present analysis to another kind of potentials, such as the
periodic potentials, one may expect interesting issues with non trivial
effects.
We also believe that our theoretical predictions could stimulate new 
experimental work in the SO and Rabi coupled BECs research.
Another interesting question, such as the study of BECs with SO and Rabi
couplings with higher pseudo-spin states in reduced dimensions, remain to
be investigated in the future.

\newcommand{\noopsort}[1]{} \newcommand{\printfirst}[2]{#1}
\newcommand{\singleletter}[1]{#1} \newcommand{\switchargs}[2]{#2#1}

\end{document}